\documentclass[acmsmall,authorcopy]{acmart}

\AtBeginDocument{%
  \providecommand\BibTeX{{%
    \normalfont B\kern-0.5em{\scshape i\kern-0.25em b}\kern-0.8em\TeX}}}

\setcopyright{acmcopyright}
\copyrightyear{2018}
\acmYear{2018}
\acmDOI{10.1145/1122445.1122456}

\setcopyright{none}

\usepackage{_macros}

\newcommand{\bstart}[1]{\vspace{1mm} \noindent{\textbf{#1.}}}

\newcommand{\name}{VisConductor}

\begin{document}

\title[\name{}]{\name{}: Affect-Varying Widgets for Animated Data Storytelling in Gesture-Aware Augmented Video Presentation}

\author{Temiloluwa Femi-Gege}
\email{tpfemige@uwaterloo.ca}
\affiliation{
  \institution{University of Waterloo}
  \city{Waterloo}
  \state{Ontario}
  \country{Canada}
}

\author{Matthew Brehmer}
\email{mbrehmer@tableau.com}
\affiliation{
  \institution{Tableau Research}
  \city{Seattle}
  \state{Washington}
  \country{USA}
}

\author{Jian Zhao}
\email{jianzhao@uwaterloo.ca}
\affiliation{
  \institution{University of Waterloo}
  \city{Waterloo}
  \state{Ontario}
  \country{Canada}
}


\begin{abstract}
Augmented video presentation tools provide a natural way for presenters to interact with their content, resulting in engaging experiences for remote audiences, such as when a presenter uses hand gestures to manipulate and direct attention to visual aids overlaid on their webcam feed.
However, authoring and customizing these presentations can be challenging, particularly when presenting dynamic data visualization (i.e., animated charts). 
To this end, we introduce \name{}, an authoring and presentation tool
that equips presenters with the ability to configure gestures that control affect-varying visualization animation, foreshadow visualization transitions, direct attention to notable data points, and animate the disclosure of annotations. 
These gestures are integrated into configurable widgets, allowing presenters to trigger content transformations by executing gestures within widget boundaries, with feedback visible only to them.
Altogether, our palette of widgets provides a level of flexibility appropriate for improvisational presentations and ad-hoc content transformations, such as when responding to audience engagement.
To evaluate \name{}, we conducted two studies focusing on presenters ($N=11$) and audience members ($N=11$).
Our findings indicate that our approach taken with \name{} can facilitate interactive and engaging remote presentations with dynamic visual aids. Reflecting on our findings, we also offer insights to inform the future of augmented video presentation tools.
\end{abstract}

\ccsdesc[500]{Human-centered computing~Visualization systems and tools}
\ccsdesc[300]{Human-centered computing~Gestural input}
\ccsdesc[300]{Human-centered computing~Mixed / augmented reality}
\keywords{Visualization, Augmented Video Presentation, Gestural Input, Animation}

\begin{teaserfigure}
    \centering
    \includegraphics[width=1\textwidth]{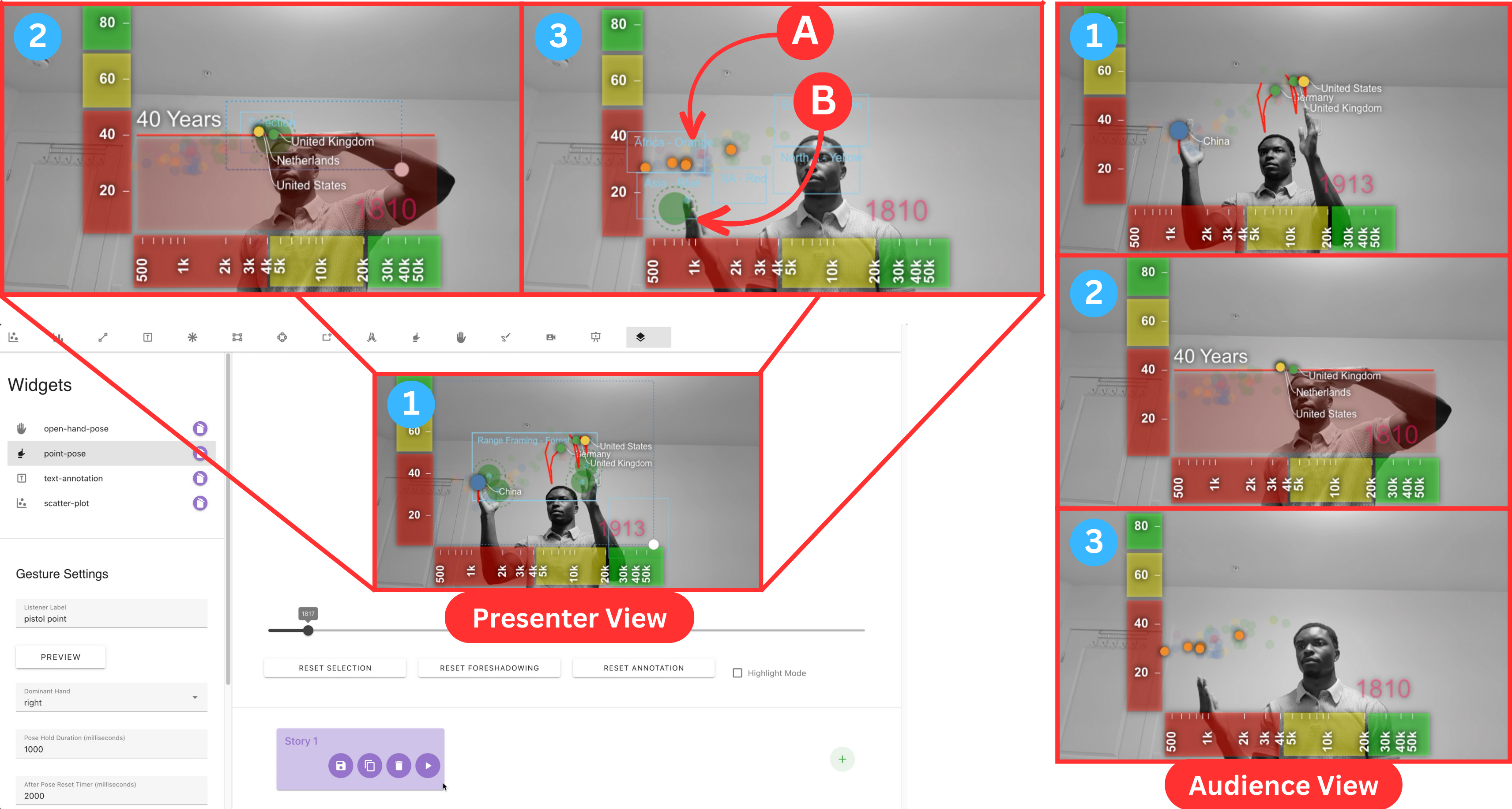}
    \vspace{-8mm}
    \caption{
        A presenter uses \name{} to deliver an augmented video presentation about life expectancy and income around the world.
        \name{} provides a local presenter view (left) as well as a remote audience view (right); the former is equipped with additional widgets to ensure precise gestural control of the overlaid visual aid (in this example, an animated scatterplot). 
        \textbf{(A):} A \textsl{Gesture Widget} outlines the boundary within which the presenter's hand gesture should be placed to trigger an animation. 
        \textbf{(B):} Upon detecting the presenter's gesture, feedback in the form of a progressively expanding green circle appears, indicating the duration for which the gesture should be maintained for accurate detection.
    }
    \vspace{5mm}
    \label{fig:presenter_vs_audience}
\end{teaserfigure}

\maketitle

\section{Introduction}

The recent surge in remote work and education has triggered an evolution of presentation tools that enhance synchronous communication with audiences at a distance~\cite{barrero_evolution_nodate, holtz_effects_nodate}, tools that convey information and accompanying visual aids in an engaging manner.
Virtual camera applications (e.g., mmhmm~\cite{mmhmm}, OBS~\cite{obs2022}), teleconference applications (e.g., WebEx \cite{webex}), and operating systems (e.g., macOS's Presenter Overlay \cite{apple2023presenter}) are now enabling speakers to easily composite their webcam video with their visual aids. 
\rev{This development means that remote audiences no longer need to divide their attention between the video of a presenter and their screen-shared visual aids.}
\rev{While screen-sharing visual aids and controlling them via mouse-based interaction will continue to remain common in informal teleconference meetings between colleagues, more formal types of presentations such as business pitches, class lectures, and conference talks are increasingly delivered online. 
It is in these contexts where communication is more effective when the audience can see the presenter and their non-verbal body language~\cite{Zeng2022GestureLensVA,Kang2016FromHT,Tieu2017CospeechGP}.}
In this paper, we build upon recent work in augmented video presentation that goes beyond visual compositing, in which a presenter's hand gestures are used to manipulate and direct attention to overlaid visual aids~\cite{Hall2022AugmentedCF,Liao2022RealityTalkRS,Saquib2019InteractiveBG}. 
The promise of gesture-aware augmented video presentation rests in the observation that gestures can be simultaneously operational and expressive: providing functional control of an interface while remaining an essential channel for non-verbal communication \cite{Tieu2017CospeechGP, Zeng2022GestureLensVA}. 
Ultimately, gestures enrich one's communicative repertoire, allowing for a more holistic and nuanced exchange of ideas and affects, and whether used intentionally or spontaneously, they add depth and dimension to remote communication.

Presentations about data are common in organizational and educational contexts, with data visualization being a common form of visual aid for information delivery~\cite{riche_data-driven_2018, fekete_value_2008}.
However, authoring and delivering gesture-aware augmented video presentations about data \rev{during a live teleconference call} is challenging~\cite{Hall2022AugmentedCF}, particularly if the content is dynamic (i.e., animated charts).
It is essential to recognize the power of animation when storytelling with data~\cite{Chu2016}, particularly as animating charts in a presentation can be critical for retaining audience attention and eliciting a range of affective responses~\cite{Lan2021KinetichartsAA}.
However, animating a chart could involve one or more distinct types of transformation: scrubbing to control temporal playback, disclosing annotations, selecting or highlighting elements, or borrowing cinematic techniques such as foreshadowing to build anticipation~\cite{Li2020ImprovingEO}.
In conventional slide-based presentation tools, presenters use
peripheral input devices (e.g., a clicker, a mouse) to trigger predefined animated transitions or motion effects.
Relying on a planned series of animation triggers limits the dynamism and spontaneity of a presentation, and in the context of data visualization, this approach precludes satisfying ad-hoc interactions with a chart, such as in response to audience questions about aspects of the data.
Returning to the prospect of controlling animated charts with gestures, we face the question of how these gestures would be specified, as well as how they would affect the audience experience.



In light of these challenges, we introduce \emph{\name{}}, a tool that allows presenters to author and perform augmented \rev{live} video presentations featuring dynamic visualization controlled via hand gestures (Figure~\ref{fig:presenter_vs_audience}). 
\name{} allows for the specification of various animated transformations to charts via a modular widget-based design, with widgets that simplify the authoring process and provide visual cues to presenters during presentation delivery.
We deliberately restricted the gestural vocabulary associated with these widgets, so as to strike a balance between operational control and communicative expressiveness.

To evaluate \name{}, we conducted two studies focusing on presenters ($N=11$) and audience members ($N=11$), respectively.
In the presenter study, we asked participants to prepare a presentation using \name{} and were given a dataset and story along with a sequence of speaking points, however, we allowed them the freedom to specify gestural widgets that would best serve them in delivering points of the story. 
In the audience study, we asked participants to watch a remote presentation given by one of the researchers. 
Our results suggest that \name{} provides an effective authoring and presenting experience for presenters as well as an engaging experience for audience members. 
To summarize, our contributions include:
\rev{
\begin{itemize}[noitemsep, topsep=0pt]
\item The design and development of \name{}, a tool for authoring and delivering live augmented video presentations featuring dynamic data visualization, which entails controlling animation playback, revealing annotations, and foreshadowing visual changes, all activated by gestures that are simultaneously operational and communicative.
\item Empirical findings from two user studies evaluating \name{}, considering the perspectives of presenters and audiences, respectively, which could inform the future design of augmented video presentation tools that allow presenters to speak about data to remote audiences during live teleconference calls. 
\end{itemize}
}

\section{Background \& Related Work}
\label{sec:relatedwork}

Our research builds on an understanding of gestural communication as well as the use of gestural interaction and data visualization in presentation contexts. 

\subsection{The Communicative Use of Gestures}
\label{sec:relatedwork:gestures}

Gestures are a multifaceted tool in communication, often conveying subtleties and nuances that spoken words cannot capture~\cite{goldin2023thinking}. 
They act as visual complements, sharpening the message's clarity and making it more accessible and memorable for the audience.
This visual support not only heightens audience engagement but also fosters better retention of the information shared.
For instance, Zeng \etal~\cite{Zeng2022GestureLensVA} documented the importance of gestures across diverse communication settings, from public speaking to daily interpersonal interaction.
Their findings, along with those of Kang \etal~\cite{Kang2016FromHT} suggest that the likelihood of effective communication is greater when the gestures used to synchronize and resonate with the spoken content.
While this synchronization often emerges naturally as a spontaneous reaction to a spoken narrative, speakers may also use gestures as a conscious strategy to emphasize and persuade.
Finally, being particularly relevant to our focus on speaking about data, Tieu \etal~\cite{Tieu2017CospeechGP} documented the amplifying effect of gestures when a speaker is presenting complex or abstract ideas.

According to taxonomies of communicative gestures~\cite{kendon_gesture_2004, Carstens2019AdviceOT}, 
\textit{deictic} gestures such as pointing direct attention, while \textit{iconic} gestures visually represent specific ideas or objects and \textit{metaphoric} gestures symbolize abstract concepts. 
\rev{Based on an analysis of 5,500 mid-air hand gestures, similarly, Aigner \etal classified gestures as follows: pointing, semaphoric, pantomimic, iconic, and manipulation \cite{aigner2012understanding}}.
In the context of presentations, Carstens~\etal~\cite{Carstens2019AdviceOT} noted that a presenter's one-handed gestures are often deictic \rev{(\eg pointing)}, while two-handed gestures tend to be iconic or metaphoric. 
In our work, we primarily respond to a presenter's deictic gestures, though we also respond to a metaphorical gesture for the passage of time-based on the motion of an analog clock.
\rev{We additionally build upon the realization that gestures can be simultaneously deictic and manipulative~\cite{Hall2022AugmentedCF}.}

\subsection{Gesture-Controlled Presentation Interfaces}
\label{sec:relatedwork:gestureux}

Presenters frequently employ communicative gestures when speaking, however in conventional presentation settings in which a presenter and their audience are co-located, these gestures are often performed at a distance from the presenter's visual aids, particularly when using a projected display, resulting in potentially ambiguous deictic references~\cite{fourney_gesturing_2010}.  
This separation is greater in most online presentations, in which a video of the presenter and a screen-shared video feed of their visual aids are shown in separate windows or panels.
Additionally, it is typical for presenters to position themselves such that their camera only captures them from the shoulders upwards, so any communicative gestures they make during their presentation go largely unseen by their audience. 
\rev{
Hauber \etal~\cite{spatialityinvideoconferencing} highlighted the importance of spatiality and social presence in videoconferencing by comparing 2D and 3D interfaces.  
Moreover, research on remote collaboration software indicates that superimposing a live video of users onto their shared content can significantly enhance collaboration~\cite{3dboard, collaboard, teleboard}. 
Part of this effectiveness can be attributed to the preservation of eye contact in addition to the visibility of gestures, which closely mimics face-to-face interaction \cite{humansensitivityeyecontact}. 
Overall, this research helps to explain the growing popularity of augmented videoconferencing applications in which visual aids are composited with a presenter's video.
}

In this work, we additionally distinguish between communicative and \textit{operational} gestures in the context of delivering a presentation, with the latter representing those deliberately designed for controlling a presentation interface. 
Cao~\etal~\cite{cao_evaluation_2005} found that hand gestures were favored relative to a laser pointer for controlling a presentation interface. 
However, Fourney~\etal~\cite{fourney_gesturing_2010} found that when granted hand gesture control of presentation materials, the range of interaction was often limited to slide navigation; people seldom used gestures to interact with slide content, and gestural interaction events were prone to unintentional triggering.
As reflected upon by Hall~\etal~\cite{Hall2022AugmentedCF}, this unintentional triggering could be a sign of interference between operational and communicative gestures.
Presenters also struggle when a system fails to provide obvious feedback in response to the gestures being performed~\cite{baudel_charade_1993}, and this is particularly pronounced when any operational gestures are separated in physical space from the visual aids they are intended to control.


Developments in real-time computer vision technologies such as Kinect \cite{shotton2013real}, OpenPose \cite{cao_openpose_2019}, and MediaPipe \cite{mediapipe2020} have recently yielded a collection of gesture-aware augmented video presentation tools that grant presenters functional control over visual aids composited over their video feed~\cite{Hall2022AugmentedCF, liu2023visual, Liao2022RealityTalkRS, Saquib2019InteractiveBG}. 
Collectively, these tools demonstrate the potential of using gestural control that is spatially aligned with visual aids, thereby also serving a communicative purpose in that the audience's attention is no longer divided between a speaker and their content.
However, even with these augmented video presentation tools, presenters still struggle to remember which gestures to perform at different points in a presentation, or to anticipate the effects that gestures will have on their visual aids~\cite{Saquib2019InteractiveBG}.
\name{} aims to address these challenges with gesture-controlled presentation tools with an authoring interface comprised of modular \textit{Gesture Widgets}, along with
visual feedback provided to the presenter as they execute these gestures at performance time.  

\subsection{Data Visualization in Presentation Contexts}
\label{sec:relatedwork:visualization}

Visualization is instrumental in translating data observations into impactful narratives~\cite{heer_interactive_nodate, kosara_storytelling_2013, satyanarayan2014authoring, segel_narrative_2010}, with considerations that go beyond mere choices of representation (\ie,~visual encoding). 

The first major consideration relevant to our work is the strategic use of animation for enhancing audience engagement and comprehension~\cite{robertson_effectiveness_2008}.
While animated visualization can effectively demonstrate changes over time or evolving patterns, it can also contribute to the affective aspect of a narrative about data~\cite{amini_hooked_2018,Lan2021KinetichartsAA}. 
This emotional connection is particularly vital in presentations, as it can promote audience engagement and information retention \cite{fisher_animation_nodate}. 
To this end, Lan \etal~\cite{Lan2021KinetichartsAA} identified the role of animation speed and easing functions on the affective impact of dynamic visualization. 
Similarly, Li~\etal \cite{Li2020ImprovingEO} demonstrated ways to employ visual foreshadowing in anticipation of animating a chart as a way to direct audience attention and keep them engaged.

While animation control can be realized using many data visualization libraries and tools, 
specialized packages such as gganimate for R~\cite{gganimate} or interactive applications such as Data Animator~\cite{thompson_data_2021}, DataClips \cite{amini2017}, and Roslingiﬁer \cite{Shin2022RoslingifierSS} surface convenient animation specification affordances.
While we draw inspiration from the options that these specialized tools provide, their resulting specifications define automatic animation or playback triggered by conventional mouse and keyboard input, whereas we connect animation playback to real-time gestural input.

Another consideration is the where and how to employ highlighting and annotation, so as to direct the audience to the most crucial aspects of the data \cite{stolper_emerging_nodate}. 
Annotations can assume many forms and be applied to various visual entities~\cite{Ren2017ChartAccentAF}, from axes and coordinate spaces to data marks; they can directly reflect values in the form of mark stroke or fill highlights, mark labels and trend lines~\cite{subramonyam2018smartcues}, or they might serve to provide additional context not captured in the underlying data.
%

\subsection{Visualization in Augmented Video Presentations}
\label{sec:relatedwork:augchiro}

In 2010, the late public health expert Hans Rosling produced a short documentary with the BBC entitled \textit{``200 Countries, 200 Years, 4 Minutes''}~\cite{bbc2021youtube}. 
In this video, Rosling stands in an empty room, appearing in the middle of the frame. Gradually, an animated scatterplot depicting the dynamics of public health metrics over time is composited in the foreground, seemingly controlled by his gesticulations. 
The video compositing and animation were both added in post-production, though the filming did require very deliberate choreography on Rosling's part, so as to appear as though he was interacting with a gesture-aware interface. 
In doing so, Rosling motivated a line of work combining visualization in real-time gesture-aware augmented reality video, including this paper.


Hall~\etal~\cite{Hall2022AugmentedCF} recently demonstrated the detection of gestures to highlight values in composited bar, line, area, and pie charts.
However, that work stopped short of using gestures for chart animation, such as showing how values change over time.  
To this end, \name{} provides the means to specify the gestural control of dynamic representations of data, namely animated scatterplots and bar charts.
These types of dynamic visualization differ from the alternating state-based animation of diagrams appearing in a demonstration of Saquib \etal's gesture-controlled video performance system~\cite{Saquib2019InteractiveBG}; in contrast, our animations reflect values bound to a continuous time-series. 

Beyond gestural control for animating time-oriented data visualization, \name{} also provides a graphical authoring interface for specifying these gestures, incorporating familiar affordances from tools for web-based data storytelling (e.g.,~\cite{amini2017,satyanarayan2014authoring,Shin2022RoslingifierSS}), including those for communicating intended affects~\cite{Lan2021KinetichartsAA}. 
However, unlike these prior tools, \name{}'s output medium is augmented video, with animation triggered by gestural input rather than by clicks or timers.  
\section{Design Considerations}
\label{sec:considerations}



\begin{figure*}[tb]
    \centering
    \includegraphics[width=1\linewidth]{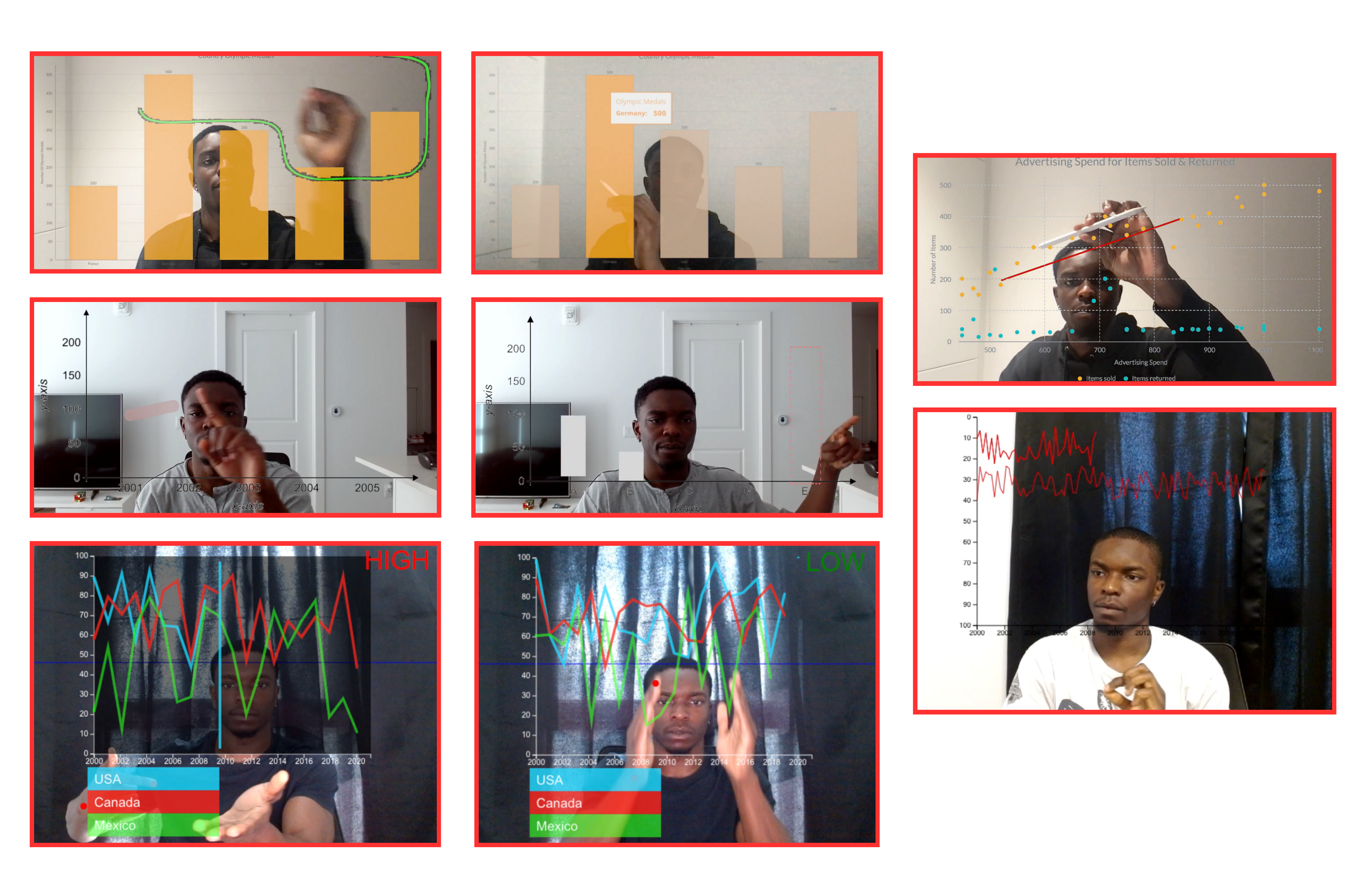}
    \vspace{-10mm}
    \caption{\rev{Frames from early low-fidelity video prototypes of animated charts composited over a presenter gesticulating.}}
    \label{fig:gesture-reel}
\end{figure*}

\rev{
Our review of prior work yielded three design considerations. 
To validate and refine these considerations,} we conducted two semi-structured interviews with a professional public speaking coach who has personally trained thousands of individuals over the course of their career~\cite{anon2023}. 
\rev{Additionally, we solicited informal feedback via unstructured interviews with two university professors and two graduate students, with each having experience giving technical presentations to remote audiences via videoconferencing applications.
In each case, we sought feedback} on our design exploration with respect to the use of gestures to control dynamic visualization in augmented video presentations.
\rev{Specifically, we solicited responses to low-fidelity video prototype demonstrations shown in Figure~\ref{fig:gesture-reel}, inspired by the prior work summarized in \autoref{sec:relatedwork:visualization}. We produced chart animations in Adobe After Effects \cite{AdobeAfterEffects} and used OBS Studio~\cite{obs2022} to composite them over a video of a presenter gesticulating, so as to simulate the interactive gestural control of chart animations.}

\bstart{D1: Recognize a small set of expressive and operational gestures} 
\rev{In response to early prototype demonstrations, the speaking coach reaffirmed the need to maintain a primary focal point of visual attention for the audience and to minimize the cognitive demand of remembering a large gestural vocabulary, echoing the findings of prior work
~\cite{cuccurullo_gestural_2012, Zeng2022GestureLensVA}.}
In other words, we require a set of gestures that are both easy to remember and easy to coordinate with speech.
Moreover, \rev{the speaking coach urged us to consider gestures that would not distract the audience}; rather they should reinforce deictic references and points of emphasis in the presenter's oration, striking a balance between being expressive and operational~\cite{Hall2022AugmentedCF}.
\rev{These low-fidelity demonstrations therefore allowed us to identify and discard candidate gestures that were purely operational, as these were deemed most likely to distract audiences.} 

\bstart{D2: Provide gestural control of visualization animation with intended affect}
Prior work has underscored the \rev{role that animation plays} in engaging presentations about data.
In this context, animation is not limited to graphical transformations~\cite{Heer2007} or showing how patterns in data change over time~\cite{robertson_effectiveness_2008}; it also encompasses the strategic reveal of annotation and highlighting~\cite{Ren2017ChartAccentAF},  the use of affective motion to elicit appropriate emotional reactions to a narrative, and the use of visual foreshadowing to set audience expectations~\cite{Li2020ImprovingEO}. 
\rev{The speaking coach particularly supported our inclusion of foreshadowing in this consideration, as this tactic prepares the audience to receive new information.}
Given this multi-faceted characterization of animation, we sought distinct gestural interactions for controlling these aspects.


\bstart{D3: Support authoring and delivery with dedicated views} 
\rev{Conversations with the speaking coach and the two professors also allowed us to refine the context of use for an augmented video presentation tool, namely professional environments in which there are numerous time pressures that demand that presentation content be prepared quickly. 
This context of use underscores the need for a predictable interface that allows for the skills developed using other tools to be more easily transferred.
This sentiment was also captured in earlier work by Brehmer and Kosara~\cite{brehmer2022}, who found that not only do those who give presentations about data desire a graphical user interface that provides interactive affordances for authoring a presentation, modifying its structure, and previewing content, they also desire a secondary display visible only to them at the time of presentation delivery, following the conventions of popular commercial presentation tools (e.g., PowerPoint, Keynote)}.
In the context of gesture-controlled augmented video presentations about data, 
we anticipate that such a secondary display could be used to provide visual feedback with respect to gesture recognition and animation state. 
\subsection{Usage Scenario}
\label{sec:considerations:scenario}

Given our three design considerations, we now describe a scenario in which a presenter uses gestures to control an augmented video presentation featuring dynamic data visualization; for a video of this scenario, please refer to our \textbf{supplemental material}.
The presenter in our scenario is John, a public health official, and his presentation is about the shifts in monetary and healthcare metrics in countries around the world over the course of the last two centuries.
We describe the authoring and delivery interface (\textbf{D3}) used by John in the next section; here, we describe the viewing experience of his remote audience.

John initially sets the scene by compositing a scatterplot over his webcam video. 
He proceeds to reveal a title with a deliberately-placed open hand gesture, as shown in Figure~\ref{fig:usage_scenario}-A.
This title text is revealed slowly, intended to mirror the profound and lingering impacts of war on public health, thereby establishing an emotional tone.
He similarly directs attention to the elements of the scatterplot via a series of open hand pointing gestures, proceeding from the axes to the data points, highlighting each of them in turn, albeit with a neutral affect. 

Next, he directs the audience's attention toward a few nations of interest. 
With pointing gestures aimed at specific data points, he triggers an animation in which these data points brighten while others dim (Figure~\ref{fig:usage_scenario}-B).
To herald changes in the data over time, John then uses a bimanual framing gesture to encircle a group of data points, highlighting them and revealing their projected trajectories (Figure~\ref{fig:usage_scenario}-C),
foreshadowing future public health trends in these countries. 

Lastly, John animates the scatterplot to illustrate changes in the data over time while simultaneously eliciting an appropriate affective response from the audience.
This animation is controlled by John's performance of a circular dialing gesture, evoking the passing of time as indicated by the movement of an analog clock.
During this animation, some of the circular marks corresponding with countries smoothly and ephemerally morph into downward-facing arrows, reinforcing the devastating impact of war (Figure~\ref{fig:usage_scenario}-D).
Once he reaches the most recent state of the data, John stops dialing to comment on the contemporary state of public health around the world.


\section{\name{} System} 
\label{sec:system}

\name{} is a browser-based application for authoring and delivering augmented video presentations about data. 
The name is an allusion to how an orchestra conductor's hand gestures direct the performance of various instruments; in our case, hand gestures control various animation effects, including but not limited to visualization playback reflecting changes in data over time. 

\subsection{Presenter Interface}
\label{sec:system:interface}

\begin{figure*}[tb]
    \centering
    \includegraphics[width=1\linewidth]{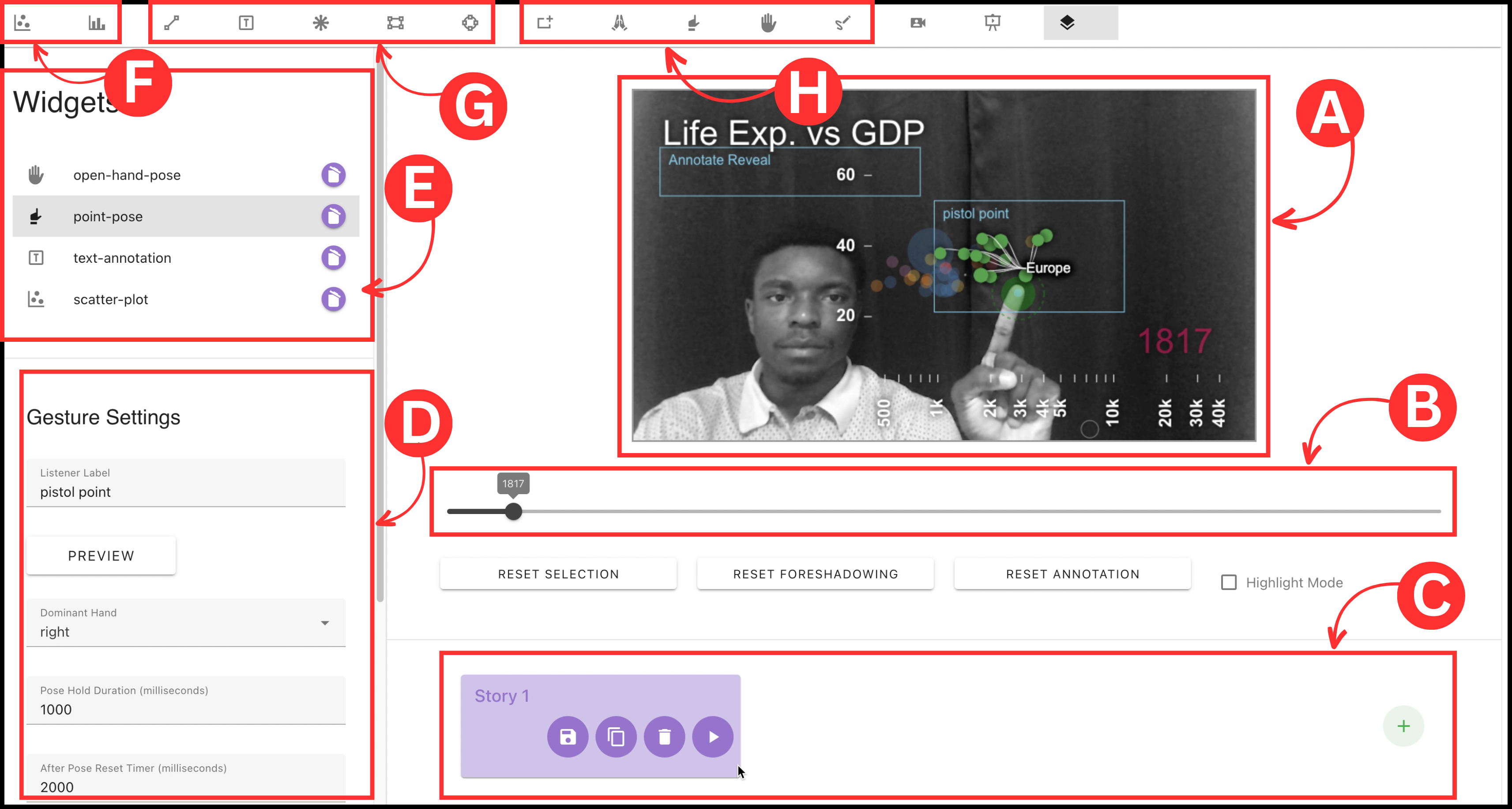}
    \vspace{-6mm}
    \caption{
        The presenter interface of \name{} consists of: 
        \textbf{(A)} the \textit{Presentation Preview}, \textbf{(B)} the \textit{Timeline Slider}, \textbf{(C)} the \textit{Storyline Tab}, \textbf{(D)} the \textit{Widget Settings Tab}, \textbf{(E)} the \textit{Widget List}, and a palette of 
        \textbf{(F)} \textit{Chart Widgets}, \textbf{(G)} \textit{Annotation Widgets}, and \textbf{(H)} \textit{Gesture Widgets}.
    }
    \label{fig:general_layout}
\end{figure*}

The presenter interface of \name{}, shown in Figure~\ref{fig:general_layout}, consists of several components. 
First, the \textit{Presentation Preview} \textbf{(A)} provides a preview of what the remote audience sees---the presenter's webcam feed composited with charts and annotations---additionally augmented with feedback visible only to the presenter (\textbf{D3}), indicating gestural activation cues and the progress of an animation.
Next, the \textit{Timeline Slider} \textbf{(B)} indicates the progress of an animation and allows for conventional cursor-based scrubbing to specific points in an animation.
The \textit{Storyline Tab} \textbf{(C)} lists distinct segments of the presentation, which are analogous to the convention of slides in a PowerPoint or Keynote presentation. 
The \textit{Widget Settings Tab} \textbf{(D)} provides controls for specifying the parameters of widgets, while the \textit{Widget List}  \textbf{(E)} lists the widgets currently in use for the selected segmented, which may include one or more \textit{Chart Widgets} \textbf{(F)}, \textit{Annotation Widgets} \textbf{(G)}, and \textit{Gesture Widgets} \textbf{(H)}.
The ordering of widgets within the \textit{Widget List} reflects an ordering of layers, adopting a convention familiar to users of tools like Photoshop or Figma; the implication of this layering is that presenters can overlay multiple widgets within a single presentation segment. 

\subsection{Chart, Gesture, \& Annotation Widgets}
\label{sec:system:widgets}

Every augmentation of a presenter's webcam video in \name{} is encapsulated as a modular widget, 
of which there are three types. 

\begin{figure}[tb]
    \centering
    \includegraphics[width=1\linewidth]{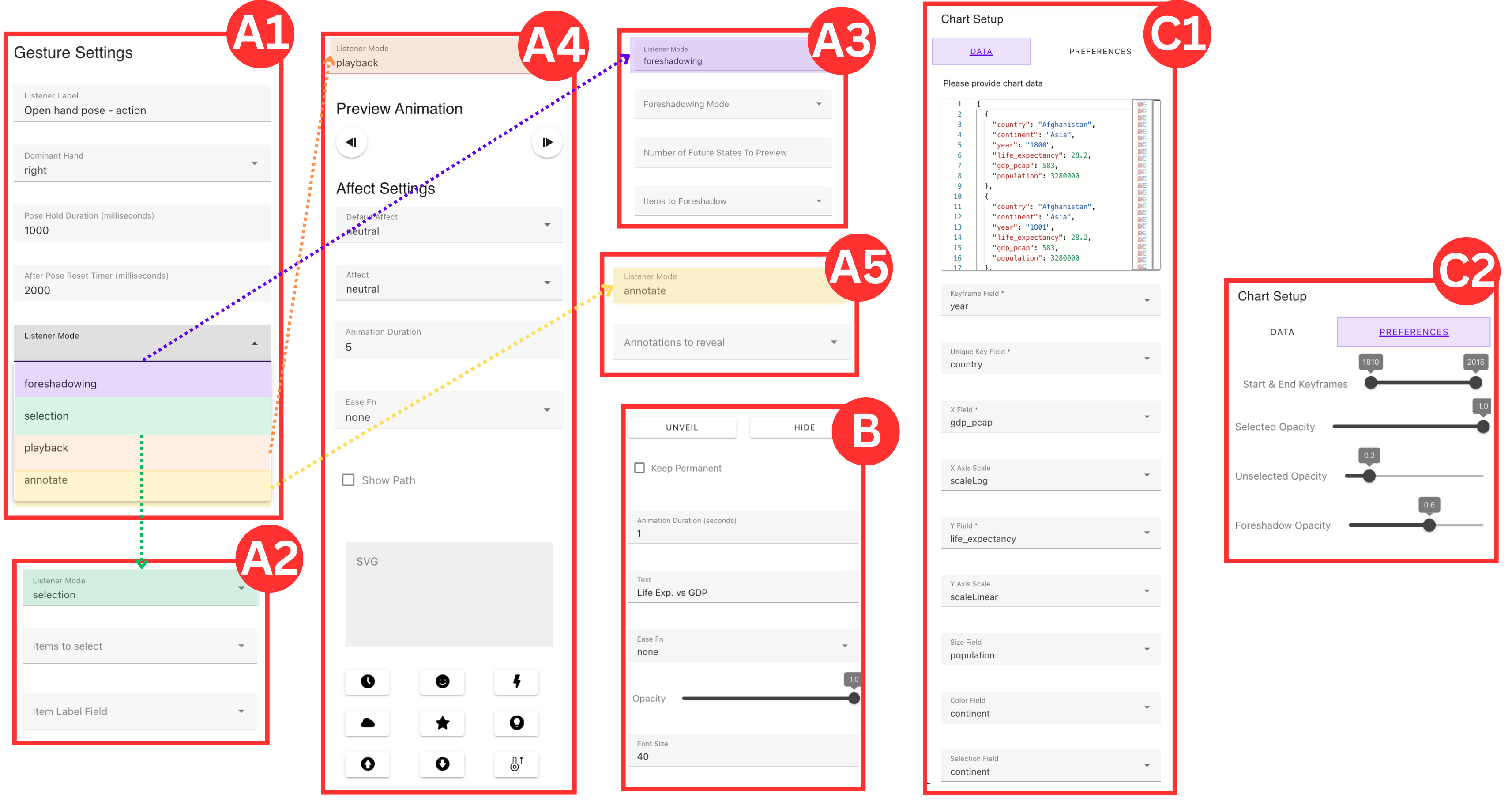}
    \vspace{-7mm}
    \caption{An overview of widget parametrization in \name{}: 
    \textbf{(A1)} \textit{Gesture Widget} settings for gesture type, recognition duration, which hand to recognize for one-handed gestures, and which operation the gesture will trigger: \textbf{(A2)} \textit{Selection}, \textbf{(A3)} \textit{Foreshadowing}, \textbf{(A4)} \textit{Playback}, or \textbf{(A5)} \textit{Annotation};
    \textbf{(B)} \textit{Annotation Widget} settings include those for specifying annotation type, color, opacity, reveal duration, and reveal easing; 
    \textbf{(C1)} \textit{Chart Widget} settings include those for ingesting tabular data and binding fields to position, size, and color scales used in the chart, while  
    \textbf{(C2)} \textit{Chart Widget} animation preferences allow for the selection of keyframes, mark opacities, and foreshadowing design. 
    }
    \label{fig:widget_settings}
\end{figure}

First, a \textit{Chart Widget} (Figure~\ref{fig:general_layout}-F) overlays a chart over the video feed.
Currently, \name{} supports two types of charts that are frequently used to tell stories about data over time (\textbf{D2}): an animated scatterplot and a bar chart race.
In both cases, the position and size of data-bound graphical marks are animated to reflect values at a particular point in time.
Figure~\ref{fig:widget_settings}-C1 illustrates the parameters available for configuring a \textit{Chart Widget}, following familiar conventions for chart authoring exemplified in existing visualization tools (i.e., loading a tabular dataset and binding position, size, and color scales to field names). 

Second, \textit{Gesture Widgets} (Figure~\ref{fig:general_layout}-H) add gestural activation to a specified region within the video frame; these could, for instance, trigger the selection and annotation of chart elements, reveal visual foreshadowing, or initiate the playback of animations corresponding to changes in data over time (\textbf{D1}).
The parameters of a \textit{Gesture Widget} (Figure~\ref{fig:widget_settings}-A1:A5) include those for binding a gesture to an existing \textit{Chart Widget} or \textit{Annotation Widget}, as well as those for specifying recognition duration, re-detection interval, and which hand the gesture should be recognized for.

Finally, \textit{Annotation Widgets} (Figure~\ref{fig:general_layout}-G) overlay text or shapes to the video frame.
Figure~\ref{fig:widget_settings}-B shows the available parameters for specifying the visual style and revealing annotations, such as altering their color or opacity, which may have deliberate affective connotations.
While text labels for data points can be retrieved from the underlying data and revealed when binding a \textit{Gesture Widget} to a \textit{Chart Widget}, \textit{Annotation Widgets} can be used to freely place editorial text and graphical elements that do not have an explicit link to data, such as when directing attention to part of an axis or to a region within a chart where there is a conspicuous absence of data marks. 





\subsection{Gestural Vocabulary}
\label{sec:system:gestures}

\begin{figure}[tb]
    \centering
    \includegraphics[width=1\linewidth]{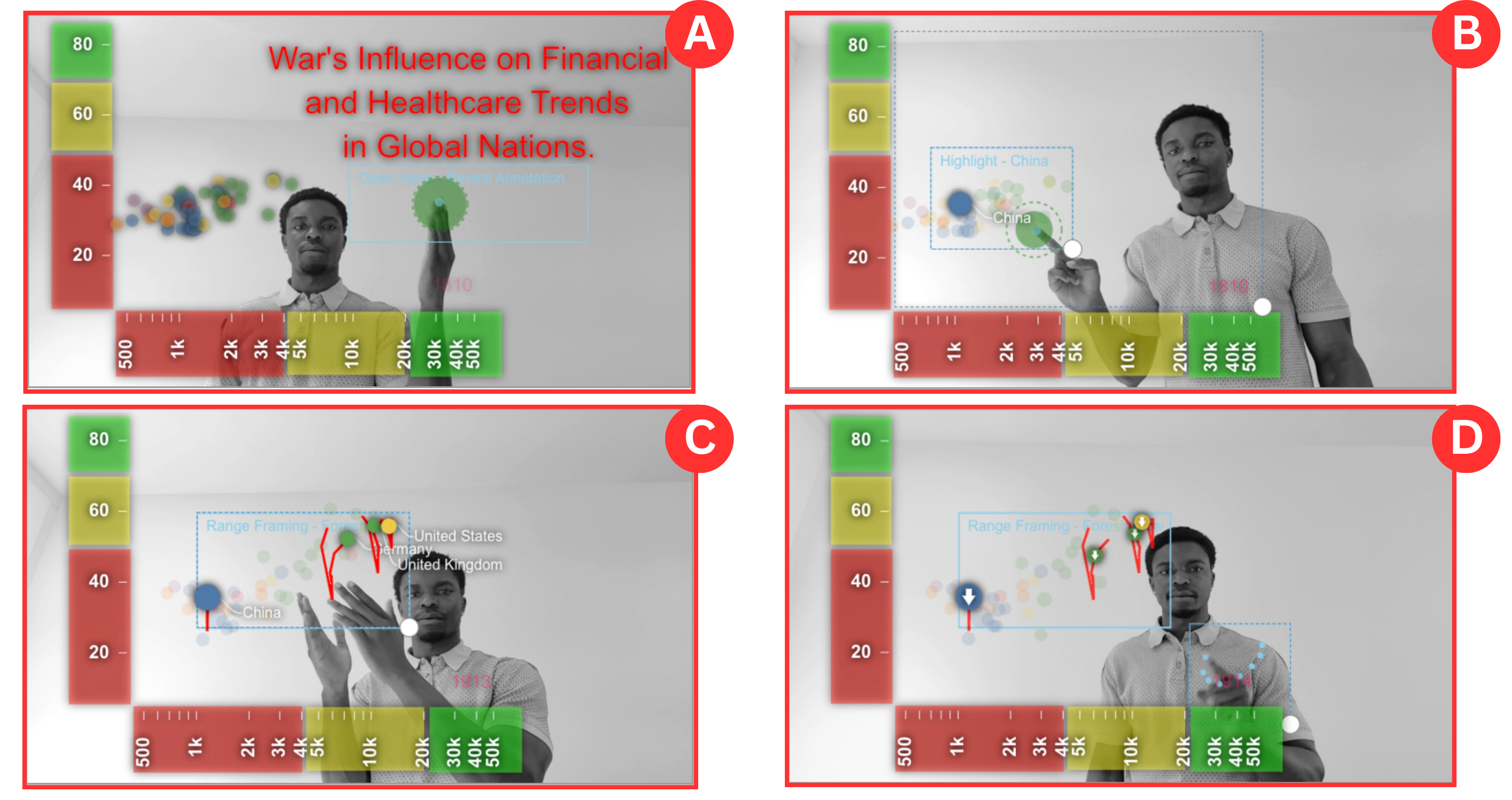}
    \vspace{-7mm}
    \caption{
        Four moments in a presentation as seen from the Presenter Preview, exemplify visual feedback for four types of gesture:
        \textbf{(A)} an \textit{Open Hand} gesture reveals a text annotation; \textbf{(B)} a \textit{Pointing} gesture highlights specific data marks within the scatterplot; \textbf{(C)} a \textit{Rectangular Framing} gesture foreshadows the trajectories of data marks appearing within that enclosed region; and \textbf{(D)} a continuous \textit{Dialling} gesture controls the playback of an animation corresponding to changes in the data over time.
    }
    \label{fig:usage_scenario}
\end{figure}

In accordance with \textbf{D1}, \name{} supports a small set of gestures that are simultaneously expressive and operational, differentiated according to whether they are single-handed or bimanual, stationary or dynamic, and which fingers are extended.  

A stationary single-handed gesture can be used to draw attention to specific visual elements.
Following Carstens' finding~\cite{Carstens2019AdviceOT} that two gestures from this category are particularly common when making deictic references in presentations, we recognize an \textit{Open Hand} gesture as well as a more directed \textit{Pointing} gesture using the index finger.
A widget for either of these gestures can be bound to an existing \textit{Annotation Widget} or \textit{Chart Widget}; Figure~\ref{fig:usage_scenario}-A shows an example of using an \textit{Open Hand} gesture to reveal a text annotation is shown in, while Figure~\ref{fig:usage_scenario}-B shows an example of using a \textit{Pointing} gesture to highlight data marks in a scatterplot.

Bimanual shaping gestures draw attention to ranges and boundaries~\cite{Shin2022RoslingifierSS}.
If the palms of the two hands are facing each other with index fingers bent, we recognize this as \textit{Rectangular Framing}, as it conveys a two-dimensional enclosure.
If the fingers are not bent, we recognize this as one-dimensional \textit{Range Framing}. 
When bound to a \textit{Chart Widget}, either type of range gesture can be used to highlight data marks appearing within a two-dimensional enclosure or a one-dimensional span, respectively. 
Figure~\ref{fig:usage_scenario}-C shows an example of a \textit{Rectangular Framing} gesture used to trigger a visual foreshadowing of trajectories for data marks originating from the enclosed area.

Rounding out our set of gestures is a dynamic \textit{Dialling} gesture inspired by radial scrolling on touch interfaces~\cite{moscovich2004,smith2004}. 
This extension of \textit{Pointing} controls the playback of animation corresponding to changes in data over time, using the visual metaphor of an analog clock (Figure~\ref{fig:usage_scenario}-D; see video).
\subsection{Combining Widgets to Communicate Intended Affects} 
\label{sec:system:affect}

\begin{figure}[tb]
    \centering
    \includegraphics[width=1\linewidth]{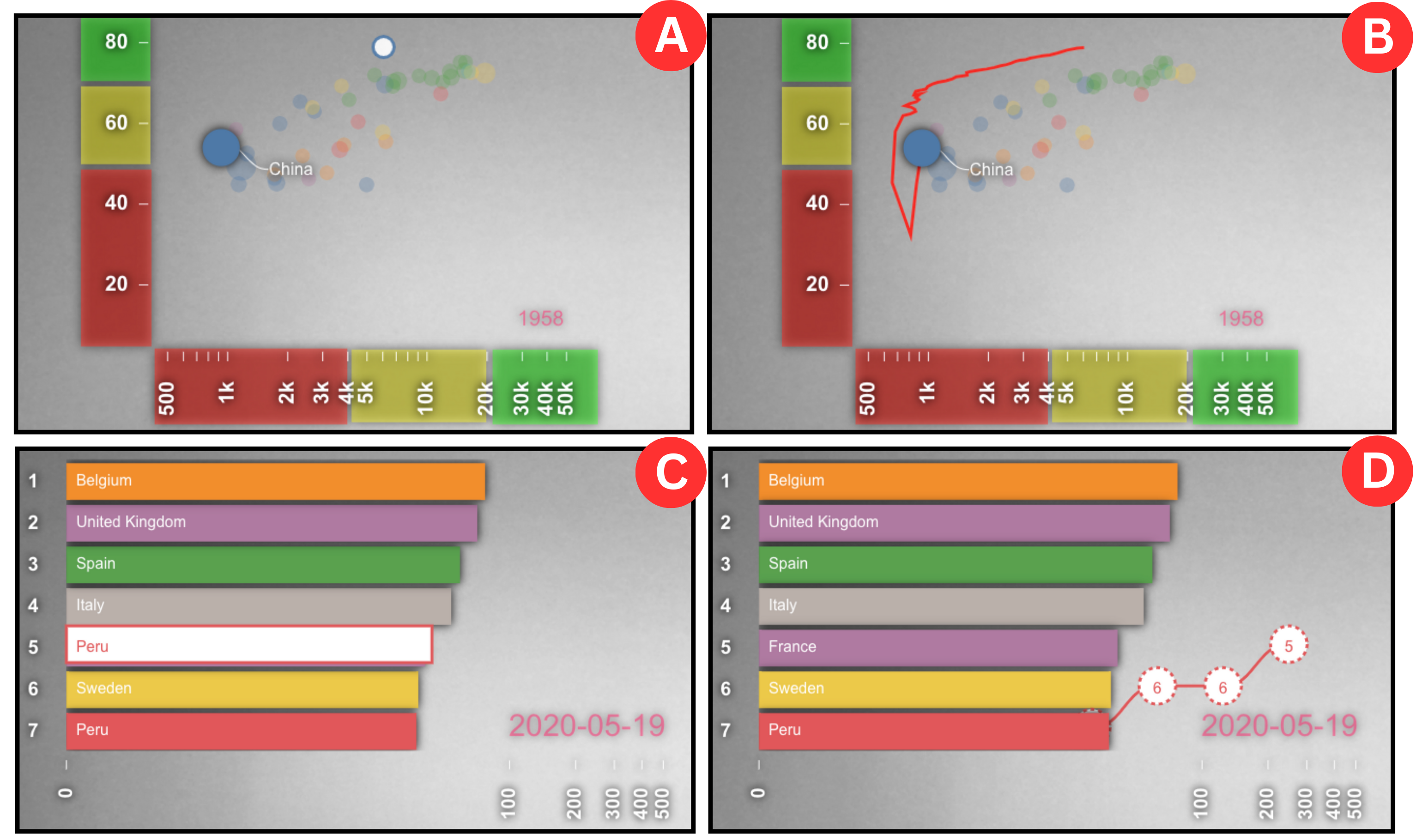}
    \vspace{-6mm}
    \caption{Visual foreshadowing applied to the two \textit{Chart Widgets}. 
        \textbf{(A)} Scatterplot position foreshadowing, showing the initial and final position of the mark(s). 
        \textbf{(B)} Scatterplot trajectory foreshadowing, showing the path of one or more data marks over time. 
        \textbf{(C)} Bar chart race position foreshadowing, highlighting the future position of a bar. 
        \textbf{(D)} Bar chart race trajectory foreshadowing, showing an ephemeral bump chart of the bar's future positions; e.g., Peru's rise from 6th to 5th place over three time steps.
    }
    \label{fig:foreshadowing_states}
\end{figure}


The \textit{Playback} settings of a \textit{Gesture Widget} (Figure~\ref{fig:widget_settings}-A4) allow presenters to establish different affects via animation (\textbf{D2}). 
In particular, easing controls provide a channel through which to convey a positive, neutral, or negative affect.
Additionally, presenters can opt to ephemerally morph selected data mark shapes into semantically-suggestive SVG-based icons during playback. 
For example, in Figure~\ref{fig:usage_scenario}-D and in the supplemental video, the presenter specifies an animation in which a scatterplot's data marks move downwards, reflecting the devastating loss of lives during World War I; next, they quickly return upwards, representing the economic recoveries that followed the war. 
The former movement is deliberately slow, while the latter is quick and bouncy, suggesting first a profound loss followed by a hopeful rebound. 
To further establish a negative affect, the circular data marks ephemerally morph into a downward arrow shape as the marks descend. 

Controls for specifying \textit{Foreshadowing} (Figure~\ref{fig:widget_settings}-A3) also provide an opportunity to elicit an appropriate affective response from the audience. 
Point foreshadowing juxtaposes a data mark's current state with a future one via an ephemeral  pulsing animation, as illustrated in Figure~\ref{fig:foreshadowing_states}-A and Figure~\ref{fig:foreshadowing_states}-C. 
In contrast, trajectory foreshadowing ephemerally sketches out path that a data mark will take in a subsequent animation, with subtle variations in the appearance of this trajectory based on the type of \textit{Chart Widget} it is bound to (Figure~\ref{fig:foreshadowing_states}). 
In both cases, visual foreshadowing is intended to build anticipation for what is soon to be revealed via animation.
\subsection{Presentation Delivery Experience}
\label{sec:system:delivery}

Like popular commercial slide presentation tools, \name{} provides both an \emph{audience view} and an \emph{presenter view} (\textbf{D3}, Figure~\ref{fig:presenter_vs_audience}).
The former can be shared with a remote audience either via the window sharing functionality offered by teleconferencing applications or via a virtual webcam application (\eg OBS~\cite{obs2022}). 
We enrich the latter with labelled blue bounding boxes for \textit{Gesture Widgets}, visible only to the presenter, delineating the boundaries of gestural activation; we show additional examples of this visual feedback in Figure~\ref{fig:usage_scenario}.
Additionally, when performing a gesture within these bounds, we show expanding green bubbles in the presenter view, ephemerally indicating the duration required for gesture recognition.

\subsection{Implementation}
\label{sec:system:implementation}

\begin{figure}[tb]
    \centering
    \includegraphics[width=1\linewidth]{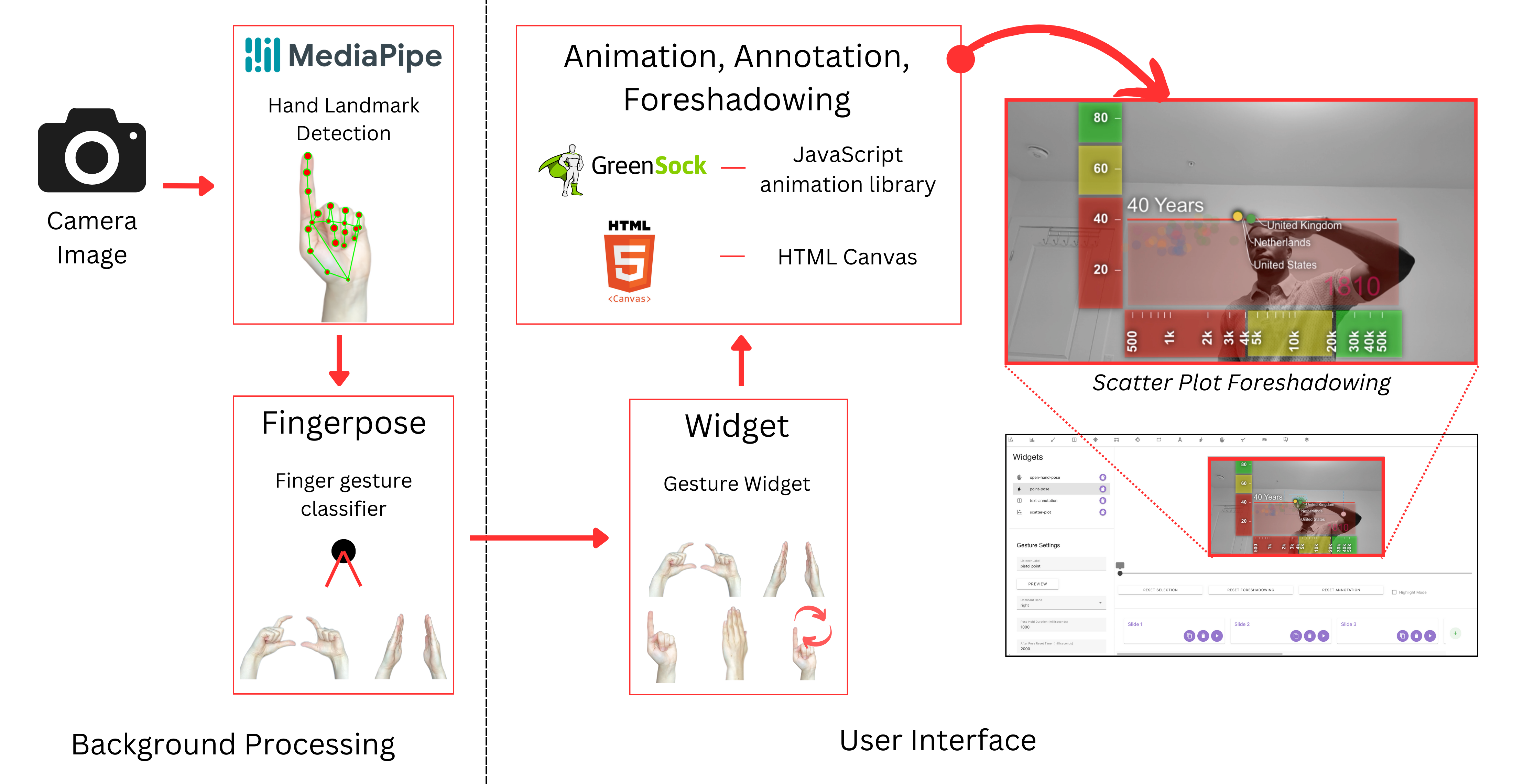}
    \vspace{-6mm}
    \caption{
    The implementation of \name{}: MediaPipe detects the presenter's hand and finger positions, which are processed by Fingerpose to classify gestures that activate \textit{Gesture Widgets}. 
    These in turn trigger the pre-configured GreenSock-based animation, annotation, and visual foreshadowing in linked \textit{Chart Widgets} and \textit{Annotation Widgets}.
    }
    \label{fig:system-diagram}
\end{figure}

\name{} is a browser-based application built using Vue.js. 
As illustrated in Figure \ref{fig:system-diagram}, the presenter's webcam video is ingested by MediaPipe~\cite{mediapipe2020}, which captures the presenter's hand and finger positions.
Fingerpose~\cite{fingerpose} uses these positions to detect predefined static and dynamic gestures.
Depending on how the presenter has assigned \textit{Gesture Widgets} to \textit{Chart Widgets} and \textit{Annotation Widgets}, these gestures will trigger selection, annotation, animation, or visual foreshadowing.
Finally, we implemented our chart animations using GreenSock (GSAP)~\cite{GSAP}.
The source code of \name{} can be found at [ANONYMIZED].
\section{Evaluation} 
\label{sec:evaluation}

We qualitatively evaluated \name{} from two perspectives: 
our first study focused on a presenter's experience of authoring and presenting using the tool, while our second study considered the audience's experience of watching a presentation delivered via \name{}.
A precedent for this methodology is Hall \etal's \cite{Hall2022AugmentedCF} evaluation of their own gesture-aware augmented video environment for giving presentations about data.
While a comparative evaluation between \name{} and their environment might have been technically feasible, we opted against this given \name{}'s focus on animation and affect, the different chart types supported by both tools, as well as the latter's absence of both an interactive authoring interface and publicly-available source code.
Hall \etal additionally compared the audience experience of gesture-aware augmented video against the convention of screen-sharing content when presenting data, finding that audiences were more engaged by the former; this finding obviated the need for us to perform a similar comparison. 
\rev{The study protocol was reviewed and approved by our institutional research ethics office.}

\subsection{Presenter Study} 
\label{sec:evaluation:presenter}

Our focus in this study was on the \textit{utility} and \textit{usability} of \name{} with respect to presentation authoring and delivery.

\bstart{Participants}
We recruited 11 participants (3F, 8M) via mailing lists and word of mouth, ranging in age between 18 and 34.
Their professional backgrounds included engineering, business, and computer science; eight held post-graduate degrees and three were college graduates. 
Regarding technological proficiency with presentation applications, five self-identified as \textit{Advanced}, four as \textit{Expert}, and two as \textit{Intermediate}.
Four reported having \textit{Advanced} presentation skills, while seven reported having \textit{Intermediate}-level skills.
All participants reported experience giving in-person presentations, while three were familiar with giving online presentations.
As for the frequency of delivering presentations, eight reported giving about one presentation a month and three reported doing so weekly.
Their presentations incorporated various elements, including data visualization, text, images, diagrams, and multimedia elements such as videos.


\bstart{Procedure}
After signing a consent form and filling out a questionnaire regarding their backgrounds,
we gave the participants a walkthrough of \name{}, highlighting its features and encouraging questions.
Next, we introduced them to a template presentation with five stories \rev{where each story in the presentation contained approximately one-minute worth of content to present}, each seeded with an initial \textit{Chart Widget}: five participants encountered the bar chart race \rev{presenting data on COVID deaths per 100,000 people across various countries between 2020 and 2022}, while six experienced the animated scatterplot \rev{showing the relationship between countries' GDP per capita and average life expectancy}.
Next, participants proceeded to add and configure additional \textit{Chart}, \textit{Annotation}, and \textit{Gesture Widgets} at their own pace, 
though we provided assistance when requested.
We then asked participants to prepare a presentation using the tool based on a provided script of speaking points.
Lastly, we conducted a semi-structured interview about their experience.
The whole study lasted about one hour and each participant received \$20.
\subsection{Audience Study} 
\label{sec:evaluation:audience}

The primary objective of this study was to assess the tool's efficacy in presenting visualization with affect-varying animations, visual foreshadowing, and highlighting, with a specific focus on how \name{} might affect audience engagement.

\bstart{Participants}
We recruited another 11 participants (4F, 7M), ranging in age between 18 and 44, having professional backgrounds that included engineering, computer science, and business. 
Five held post-graduate degrees and four were college students, while the remaining two had completed college. 
When assessing their data visualization literacy, six self-identified as \textit{Advanced}, four as \textit{Intermediate}, and one as a \textit{Beginner}.
Most (nine) frequently attended presentations on a daily (1), weekly (7), or monthly (1) basis.
All of the participants had attended in-person presentations, while six had attended online presentations.


\bstart{Procedure}
After signing the consent form, participants completed a questionnaire about their past experiences.
Then, participants watched a three-minute presentation conducted in \name{} delivered via Zoom.
While the presentation content varied (five saw a presentation featuring a bar chart race while six saw one featuring an animated scatterplot), the presentation featured all of \name{}'s major features.
After watching the presentation, participants were asked to summarize the content, which helped us gauge if they understood the presentation content and whether they had paid sufficient attention.
Finally, we conducted a semi-structured interview and asked participants to complete a questionnaire regarding their experience.
Each study session lasted about 30 minutes, and each participant received \$15.

\subsection{Observations and Participants' Reflections} 
\label{sec:evaluation:findings}
Following an iterative qualitative analysis approach, we categorized findings from both studies according to common themes. Throughout this section, ``P\#'' denotes a presenter study participant while ``A\#'' denotes an audience study participant.

\bstart{General utility} 
Presenters appreciated having gestural control of a presentation.
P7 likened the experience to that of performing, stating that \qt{it gives the idea that I have more control over my slides \ldots~almost like a performance for the audience.}
Further, P2 suggested that such a tool would make presenters commit more when giving a presentation: \qt{Working and doing physical gestures rather than just talking through the slides \ldots~it might be engaging for them as well,} a reflection that engagement is not only a concern for a remote audience. 
A7 specifically lamented the relative difficulty of animating charts in other tools, asking 
\qt{how much effort does this take to set up?  \ldots~What level of abstraction do I deal with or is it video editing?}
One potential answer is a statement by P10, who spoke to \name{}'s capacity for configuring animation via gestural interaction in contrast to how animations are specified in other platforms: \qt{The ease of interaction is what's the best point~\ldots~for [a] simple presentation tool like PowerPoint, it is harder to go within the presentation itself to point to stuff, play or annotate stuff~\ldots~it takes much more technical expertise.}


\bstart{Effectiveness of animation for communicating affect}
A9 spoke about the benefit of controlling an audience's visual attention, implying that with \name{}, \qt{you can control what you want the audience to see~\ldots~especially if you try to convey something emotional,}
suggesting its potential for affective resonance.
Meanwhile, A10 took specific note of the animations that ephemerally morphed data marks into icons that conveyed semantic and affective resonance:
\qt{There was like an arrow or something that showed up when I was expecting the circle to drop~\ldots~I feel like these small things really help me understand the point you're trying to convey.} 
A9 also remarked that these animations, particularly when spatially superimposed with a presenter's video, can reinforce the affective nature of their gestures:
\qt{It helps if you want people to see your emotional gesture~\ldots~[it] makes it more clear to show which part of the presentation you want people to focus on, whether it’s the graphics or is it both you and the graphics.} 

\bstart{Effectiveness of visual foreshadowing}
While visual foreshadowing in narrative visualization has been studied independently~\cite{Li2020ImprovingEO}, A7 remarked upon the effectiveness of this technique while also being able to see a presenter's body language: \qt{Your face being behind the data, or being with the data, definitely emphasizes the importance of where to look~\ldots~especially the World War One part where you show the trend lines before the actual animation plays out~\ldots~that was super interesting.}
From a presenter’s perspective, P4 felt that foreshadowing helped both to highlight parts of a chart graph and to pique the audience's curiosity, or as P3 put it, to 
keep the audience \qt{on their toes.} 

\bstart{Effectiveness of annotation disclosure} 
A7 made specific mention of the interplay between a presenter's gestures and the visual design of annotations, such as their color and placement: \qt{Your gestures explaining the axes and the different colors and how they move~\ldots~it was easy to understand~\ldots~I don't think there were any gaps or lack of understanding because of too much information on the plot itself.}
From a presenter's perspective, P1 spoke about this interplay as well: \qt{The gestures are good, I can just highlight which position of the graph I want to tell to the audience and we can set beforehand which part I will show so it will highlight accordingly.} 
P4 further attested to the utility of connecting gestures to the reveal of annotations, noting the synergistic effect of selection and foreshadowing in emphasizing specific chart elements: \qt{[it] helps me highlight certain parts of the graph, just to show whatever is coming next.}
Presenters also discussed the relative advantages of using gestures to reveal annotations over conventional presentation tools, with 
P10 stating that \qt{in a simple presentation tool like PowerPoint, it is harder to go within the presentation itself to point to annotate stuff.}

\begin{figure}[tb]
    \centering
    \includegraphics[width=1\linewidth]{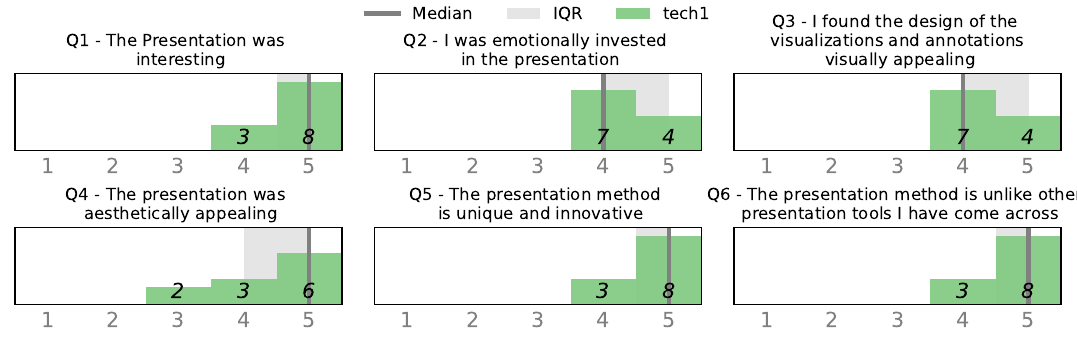}
    \vspace{-8mm}
    \caption{
        Distribution of responses to User Engagement Scale~\cite{obrien_practical_2018} questions among participants in the audience study (1 = `strongly disagree',  5 = `strongly agree'). 
        The dashed line indicates the median and the shaded region denotes the interquartile range (IQR). 
    }
    \label{fig:ues_results}
\end{figure}

\bstart{Effectiveness with respect to audience engagement}
Figure~\ref{fig:ues_results} shows the audience study participant responses to User Engagement Scale~\cite{obrien_practical_2018} questions.
The predominant sentiment was positive, reflected in a median score of 4 or above across all questions, complementing the preceding remarks about specific aspects of the presentation experience.
Overall, participants found the presentation they saw to be interesting (Q1) to the point of eliciting emotional investment (Q2), visually or aesthetically appealing (Q3-Q4), unique or innovative (Q5), and unlike presentations they had seen delivered using other tools or approaches (Q6). 
\bstart{Customizability and usability} 
P3 remarked that \qt{authoring is pretty easy~\ldots~it's really accommodative~\ldots~there are a lot of options to choose from.} 
More specifically, P9 appreciated the ability to morph chart elements into various SVGs: \qt{all the visual effects are really good~\ldots we can customize the icons~\ldots 
the tool's customizability allows presenters to create diverse and engaging animations that complement their data and message.}
However, while \name{} offers several animation parameters, P11 mentioned that it requires greater attention to design details: \qt{it means that you need to put more details into the design part~\ldots but once you get familiar with the whole system, it allows you to create the perfect slides.} 

With respect to \textit{Gesture Widgets} and their associated feedback, P3 was familiar with a widget-based approach based on experience with other tools:
\qt{it was easy for me as a first-time user to do it really fast~\ldots It felt familiar because when we use various other apps, we see these options available as widgets.} 
Meanwhile, P1 appreciated \name{}'s flexibility with respect to gesture positioning, stating: \qt{everything is manually tunable, so I can position the boxes I can position the gestures.} 
Presenters also commented on the quality of feedback from the gestural widgets, 
with P5 finding the experience to be better than expected during presentation delivery, pointing out that \qt{I expected it to be a bit more confusing and hard to handle, but it was a really good experience, I could easily communicate, understand why it's not working or why it is working.}

\bstart{Challenges}
One issue raised by presenter study participants was a need to accommodate more spontaneous interactions during a presentation. 
For instance, P6 expressed a desire for \qt{more dynamic control over charts,} emphasizing that there should not be a necessity to extensively \qt{pre-plan things} in advance. 
We interpret such comments as a need for ways that presenters can adapt to unexpected events and varying levels of audience engagement during a presentation: be able to go off-script, ad-lib, or respond to interruptions and questions. 
This feedback underscores the need to balance planned automation and real-time manual control, to ensure that presenters can seamlessly navigate prepared material without feeling overly constrained by the tool. 
To this end, P3 suggested prompts visible only to presenters that indicate when specific gestures could be performed, allowing some gestures to be skipped. 
We acknowledge that by providing participants with a list of prepared speaking points, they might have interpreted this list to be a linear sequence, however, this was never explicitly stated as such. 

Participants in the audience study raised a different concern, namely one relating to animation and gestural coherence. 
For instance, \name{} allows \textit{Gesture Widgets} to be freely placed, meaning that a misalignment of a pointing gesture with an animated referent, despite being valid from an implementation perspective, could confuse the audience and divide their attention.
This potential disconnect, according to A8, could occur if gestures do \qt{not match the visual effects or properly communicate the message,} underscoring the need for coherent gesture-animation interactions, such as by enforcing position and / or distance thresholds when linking a \textit{Gesture Widget} to an existing \textit{Chart} or \textit{Annotation Widget}. 
\section{Discussion} 
\label{sec:discussion}
We now reflect on the design of \name{} and on findings from our evaluation with reference to design requirements \textbf{D1}---\textbf{D3}.

\subsection{Design Implications for Gesture-Aware Augmented Video Communication} 
\label{sec:discussion:implications}

We pursued a modular widget-based approach in \name{}, not only as a means to balance creative flexibility and usability, but also as a way to address the unique requirements associated with presenting data (\textbf{D2}): distinguishing between visualization and annotation and establishing affective associations with trends in the data. 
While some of our reflections below are unique to presenting data, we anticipate that aspects of our research are generally applicable to gesture-aware augmented video communication in a broader sense, particularly as our other two design considerations are not unique to presenting data, with \textbf{D1} calling for a small vocabulary of expressive and operational gestures, and \textbf{D3} calling for dedicated authoring and presentation interfaces that are distinct from what an audience sees.

\bstart{Balance expressivity and operational control while imposing spatiotemporal constraints}
Prior work~\cite{Hall2022AugmentedCF} in gesture-aware augmented video presentation
warns of a potential clash between expressive and operational gestures, where the former serves as non-verbal communication with an audience and the latter controls aspects of an interface.
In \name{}, we opted for a small vocabulary of gestures that satisfy both criteria simultaneously (\textbf{D1}). 
Most of our gestures are deictic in nature, in which an extended finger or enclosed pair of hands are placed near a visual referent that will be in some way transformed, either animated, annotated, or contrasted with its future state via visual foreshadowing.
A notable exception is the \textit{Dialling} gesture, a metaphorical gesture for advancing in time; this gesture has no immediate deictic referent, though this absence of visual correspondence is appropriate here, as time is not represented visually, as it would be in the horizontal axis of a line chart, for example.
Otherwise, the spatial and temporal correspondence between a gesture and a dynamic visual aid helps to establish cause and effect and reinforces the agency of the presenter.
Considering what participants in the audience told us with respect to the reveal of annotations and visual foreshadowing, this correspondence is critical for directing the audience's attention and building rapport. 

We acknowledge the possibility that a spatial correspondence between a deictic gesture and its visual referent(s) may not always be possible, particularly if the relative positioning of the presenter and webcam puts some visual elements out of reach. 
In these cases, we envision two potential solutions. 
First, a presenter could alternatively opt to use a physical pointing device that is similarly tracked by the system.
Second, we could extend \textit{Gesture Widgets}.
For instance, a \textit{Pointing} gesture could draw a vector emanating from the extended finger, a vector that could highlight or annotate distant targets.
Another possible extension is an adaptation of the bimanual \textit{Rectangular Framing} gesture to recognize a triangular aperture formed by two hands~\cite{xia2017writlarge}; we could visually demarcate this aperture and similarly highlight or annotate targets that fall within it.

\bstart{Convey a spectrum of affects via gesture when telling stories with data}
If the goals of data storytelling are to change viewers' opinions and beliefs, to persuade them, or to inform their decisions, establishing an appropriate affect for a narrative is critical~\cite{robbins_affective_2022}. 
Unfortunately, it can be challenging to convey semantic and affective associations with the abstract visual elements used in conventional chart types.  
As we demonstrated in \name{}, a presenter can perform gestures to animate visualization elements, so as to reflect how values change over time (\textbf{D2}), where elements could, for instance, change in size, position, shape, or color.
If we consider a linearly-paced and automatically-triggered animation to be a baseline, one suggestive of more neutral affect, the ability to parameterize the easing and pacing of an animation via gesture can suggest a broader and more nuanced range of affects.
Consider our scenario (\autoref{sec:considerations:scenario}), in which the presenter drew attention to the cataclysmic effects of World War I on global health metrics followed by a triumphant rebound, using expressive gestures to reinforce this contrast.
While prior visualization authoring tools have offered extensive options to parameterize animation (\eg, \cite{Shin2022RoslingifierSS, thompson_data_2021}), including those for selecting specific affects~\cite{Lan2021KinetichartsAA}, what is unique about \name{} is that a viewer can directly perceive this parameterization in real-time by observing a presenter's gestures.

A presenter's body language is not the only way to communicate affect when animating data-bound visual elements. 
Recall that \name{} additionally provides affordances to ephemerally morph the visual appearance of abstract shapes into semantically-charged ones, such as when we morphed circles in the scatterplot into downward-facing arrows when the values fell precipitously. 
Altogether, we argue that using an expressive gesture in tandem with the selective morphing of visual elements are complementary strategy for communicating an appropriate affect in the context of data storytelling.
Prior research has highlighted the value of using semantically-rich or iconic representations of data in terms of memorability~\cite{borkin2013makes}. While we did not empirically assess memorability in our audience study, we are optimistic based on participants' comments and ratings relating to engagement that this combination of strategies would make for a memorable presentation of data.

Beyond the broad distinction between positive and negative affects, anticipation is a specific affective state that can be elicited when telling stories, particularly through the employ of non-chronological narratives.
One promising technique for establishing this sense of anticipation in the context of data storytelling is that of visual foreshadowing as described by Li~\etal~\cite{Li2020ImprovingEO}: ephemerally juxtaposing the final and initial states of visual elements and / or their intermediary states along a visible trajectory before an animation begins. 
In \name{}, presenters invoke this technique via gesture. 
This gesture-driven foreshadowing was seen not only as a way to guide the audience's attention, but also as a way to elicit curiosity and hypotheses before the presenter could explain a trend in the data during an animation.
\rev{\bstart{Provide dedicated controls and views to support the flexible and dynamic nature of presentations}}
Prior work in augmented video communication offers alternative strategies in terms of presentation delivery support. 
Hall~\etal~\cite{Hall2022AugmentedCF} proposed a \textit{what-you-see-is-what-I-see} (WYSIWIS) approach, in which there is no ambiguity with respect to what the presenter sees relative to what the audience sees. 
However, a downside of this approach is that a presenter receives no additional visual feedback or guidance as they perform gestural interactions.
Alternatively, Liu~\etal's recent work on visual captions~\cite{liu2023visual} reveals recommended visual aids in a dedicated interface adjacent to the video feed, from which a presenter can select recommended items, which in turn are composited in the WYSIWIS video.
In \name{}, authoring controls remain accessible throughout a presentation's delivery; while a presenter can share and monitor a WYSIWIS audience view (\autoref{fig:presenter_vs_audience}-right), \name{} has a dedicated presenter view (\autoref{fig:presenter_vs_audience}-left) that additionally overlays gestural activation cues and feedback in addition to the visual aids (\textbf{D3}).
Ultimately, we argue that WYSIWIS is insufficient for gesture-aware augmented video presentations about data, especially for those involving animated charts or gestures that have precisely specified spatial or temporal constraints.
A major benefit of gesture-aware augmented video presentation is the potential for presenters to deviate from a planned linear sequence.
In conventional presentation tools, the animated reveal of visual aids on a slide follows a planned order, whereas with gestural activation, these reveals can be performed in different sequences at presentation delivery time, accommodating unforeseen events such as responding to questions or interruptions from the audience. 
While this non-linearity was realized in Hall~\etal's~presentation environment~\cite{Hall2022AugmentedCF}, their gestural vocabulary and activation parameters could not be changed at presentation delivery time.
In contrast, \name{}~provides presenters with full authoring control during a presentation, allowing them to modify existing widgets or add new ones on the fly.
While we did not directly consider this scenario in our presenter study, we expect that having this interface on hand would reassure presenters.
One promising approach would be to recommend new widgets during presentation delivery in a manner similar to that of Liu~\etal's visual captions~\cite{liu2023visual}, a generative approach that would require a real-time transcription of the presenter's monologue along with any audience questions or comments. 
For example, if a presenter mentions a particular entity in the data, the system could recommend a \textit{Gesture Widget} tailored for highlighting the corresponding visual element.
However, these widget recommendations would need to be delivered in an unobtrusive manner, such as by providing affordances to disable or dismiss recommendations during a rehearsed part of a presentation and enable them during unscripted segments, such as when fielding questions from the audience.   
\subsection{Limitations and Future Work} 
\label{sec:discussion:limitations}

As a research prototype, we constrained the scope of \name{} to two types of animated chart commonly encountered in data storytelling: the animated scatterplot has been a topic of study in the visualization research community for nearly two decades~\cite{robertson_effectiveness_2008}, while the animated bar chart race is a newer convention, one recently considered in Li~\etal's work on visual foreshadowing~\cite{Li2020ImprovingEO}.
However, there are many other animated chart design conventions used in practice, from dynamic node-link representations of networks changing over time to various forms of animated thematic maps, not to mention the opportunities for animating bespoke works of visualization design. 
We acknowledge the possibility that some of our proposed \textit{Gesture Widgets} may generalize in the context of other visualization design idioms, while others may require adaptation depending on the specific visual encoding design choices in use. 
That being said, a promising potential contribution of future research would be a systematic investigation of the efficacy of common expressive and operational gestures when paired with different visual encoding channels (\eg~position, orientation, area, hue, etc.) and animated transitions~\cite{Heer2007}, as opposed to specifying unique gestures for each type of chart.
In the meantime, should \name{} evolve into a publicly-available presentation tool, its adoption would depend on the addition of affordances for using other types of visual aids in presentations, from images and text to static (\ie, non-animated) charts.  

It is also important to acknowledge the known difficulties of evaluating narrative visualization authoring tools~\cite{satyanarayan2019critical} such as \name{}.
\rev{Despite these challenges and a modest number of participants in both of our studies ($N=11$), we were nevertheless able to gather insightful quantitative and qualitative data.} 
In the presenter study, we provided participants with a template presentation featuring several widgets, along with data and speaking points to seed their own presentation.
While template reproduction or extension as well as open-ended exploration are commonly associated with evaluating visualization authoring tools~\cite{ren2018reflecting}, particularly when working with familiar constraints (\ie participant time), these activities can only provide initial assessments of utility and usability.
\rev{One aspect of usability that we did not assess involves the potential fatigue associated with performing gestures over the course of long presentations. 
However, we deem it unlikely that animated charts would be required throughout a long presentation; instead, we expect that they would be interleaved with other visual aids such as text or images that require no overt gestural interaction.}
A longitudinal deployment study of \name{} that allows participants to use their own data at their own pace would certainly be illuminating, though \name{}'s functionality would need to be extended as described above to attract prospective early adopters.

In our audience study, we could not guarantee that participants \rev{had a preexisting interest} in the topic of the presentations they saw, though they were able to provide us with a preliminary sense of whether presentations delivered using \name{} could be engaging and memorable. 
\rev{Similarly, a brief study context could not shed light on whether presentations delivered via \name{} might be used to combat so-called \textit{Zoom Exhaustion and Fatigue}~\cite{Fauville2021ZoomE}, a condition identified in remote learning environments that appears to be exacerbated by a lack of non-verbal cues.}
One possible next step would be to complement our initial results by partnering with post-secondary educators conducting online lectures in STEM fields where it is common to use charts and animations as visual aids.
\rev{Across a diversity of presentation topics and durations,} observing how students engage with lecturers and speaking to them about their experiences would be informative, particularly if the students are intrinsically motivated to learn about the subject matter being presented.

\section{Conclusion}

We introduced \name{}, an augmented video presentation tool that combines the innate communicative power of hand gestures with spatially-superimposed animated data visualization.
Recognizing that many gestures can be both communicative and operational, \name{} incorporates gestural recognition to trigger animation playback, visual foreshadowing, and the reveal of annotation.
Our evaluation, considering the perspectives of presenters and audience alike, provided insights into the utility of this medium for telling stories with dynamic representation data.
Presenters appreciated having fine-grained control over animation and annotation, while audience members were captivated by the immersive and engaging nature of presentations delivered using \name{}.
As remote communication needs and use cases continue to proliferate, tools like \name{} are poised to make presentations more than just an exchange of information, but rather an engaging and emotionally-resonant experience.




\bibliographystyle{ACM-Reference-Format}
\bibliography{references.bib}

\end{document}